\documentstyle[12pt]{article}
\def\R{ {\rm R \kern -.31cm I \kern .15cm}}
\def\C{ {\rm C \kern -.15cm \vrule width.5pt \kern .12cm}}
\def\Z{ {\rm Z \kern -.27cm \angle \kern .02cm}}
\def\N{ {\rm N \kern -.26cm \vrule width.4pt \kern .10cm}}
\def\1{{\rm 1\mskip-4.5mu l} }
\def\lsim{\raise0.3ex\hbox{$<$\kern-0.75em\raise-1.1ex\hbox{$\sim$}}}
\def\gsim{\raise0.3ex\hbox{$>$\kern-0.75em\raise-1.1ex\hbox{$\sim$}}}
\def\noi{\noindent}

\def\beq{\begin{equation}}   \def\eeq{\end{equation}}
\def\bea{\begin{eqnarray}}  \def\eea{\end{eqnarray}}
\def\nn{\nonumber}
\def\noi{\noindent}
\def\beeq{\begin{eqnarray}} \def\eeeq{\end{eqnarray}}
\newcommand\mysection{\setcounter{equation}{0}\section}
\renewcommand{\theequation}{\thesection.\arabic{equation}}
\newcounter{hran} \renewcommand{\thehran}{\thesection.\arabic{hran}}

\def\bmini{\setcounter{hran}{\value{equation}}
\refstepcounter{hran}\setcounter{equation}{0}
   \renewcommand{\theequation}{\thehran\alph{equation}}\begin{eqnarray}}

\def\bminiG#1{\setcounter{hran}{\value{equation}}
\refstepcounter{hran}\setcounter{equation}{-1}
\renewcommand{\theequation}{\thehran\alph{equation}}
\refstepcounter{equation}\label{#1}\begin{eqnarray}}

\def\emini{\end{eqnarray}\relax\setcounter{equation}{\value{hran}}\renewcommand{\theequation}{\thesection.\arabic{equation}}}

\def\frac#1#2{{#1 \over #2}}

\def\half{\ifinner {\scriptstyle {1 \over 2}}
    \else {1 \over 2} \fi}

\def\simge{\mathrel{%
    \rlap{\raise 0.511ex \hbox{$>$}}{\lower 0.511ex \hbox{$\sim$}}}}
\def\simle{\mathrel{
    \rlap{\raise 0.511ex \hbox{$<$}}{\lower 0.511ex \hbox{$\sim$}}}}

\begin{document}

\begin{titlepage}

\vspace*{0.5cm}
\begin{center}
\baselineskip=20pt
{\Large\bf Massive Yang-Mills Theory in Abelian Gauges}
\vskip1.5cm 
\end{center}
\centerline{\bf Ulrich ELLWANGER$^{\bf a}$
\footnote{Email: ellwanger@th.u-psud.fr}, 
Nicol\'as WSCHEBOR$^{\bf a,b}$
\footnote{Email: wschebor@th.u-psud.fr}}

\vskip 3 truemm

\centerline{$^{\bf a}$ Laboratoire de Physique Th\'eorique
\footnote{Unit\'e Mixte de Recherche - CNRS - UMR 8627}}  
\centerline{Universit\'e de Paris XI, B\^atiment 210, F-91405 ORSAY
Cedex, France}

\vskip 3 truemm

\centerline{$^{\bf b}$ Institutos de F\'{\i}sica} 
\centerline{Facultad de Ciencias (Calle Igu\'a 4225, esq. Mataojo)} 
\centerline{and Facultad de Ingenier\'{\i}a (C.C.30, CP 1100),
Montevideo, Uruguay}

\vskip 1cm
\centerline{\bf ABSTRACT}

We prove the perturbative renormalisability of pure SU(2) Yang-Mills
theory in the abelian gauge supplemented with mass terms.  Whereas
mass terms for the gauge fields charged under the diagonal U(1) allow
to preserve the standard form of the Slavnov-Taylor identities (but
with modified BRST variations), mass terms for the diagonal gauge
fields require the study of modified Slavnov-Taylor identities. We
comment on the renormalization group equations, which describe the
variation of the effective action with the different masses. Finite
renormalized masses for the charged gauge fields, in the limit of
vanishing bare mass terms, are possible provided a certain combination
of wave function renormalization constants vanishes sufficiently
rapidly in the infrared limit.

\vskip 2.9 truecm
\noi PACs: 11.15.-q, 11.15.Pg, 12.38.Aw
\vskip 0.5 truecm
\noi LPT Orsay 02-39 \par
\noi April 2002 \par
\end{titlepage}

\mysection{Introduction}
\hspace*{\parindent}

Nowadays it is generally accepted that massive Yang-Mills theory, in
its perturbative formulation, requires a Higgs field in order to avoid
a clash between renormalizability and (perturbative) unitarity \cite{1}. 
Nevertheless the study of massive pure Yang-Mills theories is important
for several reasons: First, perturbative calculations of many
physically interesting quantities are plagued by infrared divergences,
which require an infrared regularization. Such an infrared
regularization should be consistent, i.e. preserve renormalizability
and allow to recover perturbative unitarity in the limit where it is
removed (as long as one remains within a perturbative framework like
partonic matrix elements). Since mass terms belong to the simplest
infrared regulators, one should master their corresponding properties.
Second, lattice results on the behaviour of gauge field propagators in
the infrared regime indicate that nonperturbative effects induce
a massive behaviour of these propagators \cite{2,3}. \par

The advantage of Abelian gauges \cite{4,5} in the infrared regime of
Yang-Mills theories is that they allow to formulate most clearly the
monopole condensation picture of confinement \cite{6}. It is believed
that the confining phase of a Yang-Mills theory is dual 
to a Landau-Ginzburg or Higgs phase, in which the
off-diagonal gluons would be massive. Indeed results of lattice
simulations in abelian gauges \cite{3} indicate that here  the
propagators of the "off-diagonal" gauge fields (associated with
off-diagonal generators) and "diagonal" gauge fields (associated with
diagonal $U(1)$ subgroups) behave very differently: The massive
behaviour is observed only for the off-diagonal gauge fields \cite{3}.
Some arguments which serve to explain this phenomenon have  been
proposed in \cite{7,8}. The diagonal gauge fields (including their
monopole configurations) may then be responsable for confinement,
perhaps in some analogy to compact electrodynamics \cite{9}. Moreover
it has recently been shown that some generic families of abelian gauge
theories confine \cite{10}. We can then expect that one of these latter
models represents the long distance behavior of the Yang-Mills theory,
at energy scales below the masses of the off-diagonal gauge fields.
\par

In addition it should be noted that the non-perturbative existence of
an abelian gauge fixing for (SU(2)) Yang-Mills theory, including the
solution of the Gribov problem \cite{11}, has been proved for a
covariant, local and BRST-invariant action \cite{12}. \par

For the above reasons the study of massive Yang-Mills theory in
abelian gauges is required. As in the case of various versions of
massive Yang-Mills theories in gauges preserving the global part of the
gauge group \cite{13,14} one should be able to define a renormalizable
gauge fixing sector of the theory. This is the main task of the present
paper. As usual, to this end one has to study slightly modified
versions of Slavnov-Taylor identities and some additional symmetries
of the action, which should be preserved by the ultra-violet
regularisation (as dimensional regularisation). \par

Since we insist on a complete infrared regularisation of the theory we
study the effects of mass terms for both the off-diagonal and diagonal
gauge fields (which do not, however, have to be the same).
The required modifications of the Slavnov-Taylor identities turn out
to be quite different: Whereas mass terms for the diagonal gauge fields
require the introduction of additional sources which couple to its BRST
variation, mass terms for the off-diagonal gauge fields (accompanied by
suitable mass terms for the off-diagonal ghosts) {\it alone} can be
coped with by slight variations of the BRST transformations or
the Slavnov-Taylor identities.\par

Once renormalisability of the theory has been shown, one can proceed
and discuss particularly convenient renormalisation schemes. Since the
theory is, by construction, infrared save, one can renormalise at
vanishing external momenta. The (asymptotic) scaling behaviour of the
Green functions can be described by the Callan-Symanzik (CS) equation
\cite{15} which, strictly speaking, describes the variation of the
Green functions with the masses.

The CS equation plays a role somewhat in between the usual
renormalisation group equations in some mass independent substraction
scheme (where the first two terms of the $\beta$-functions are
universal) and the Exact Renormalisation Group (ERG) Equations
\cite{16,17}, which are based on general momentum dependent cutoff terms
(quadratic in the fields) in the action. Since also here the mass terms
can play the role of an infrared regulator (whose variation is
described by the ERGs), this issue will be discussed briefly below.\par

Given the lattice results on the massive behaviour of propagators of
the off-diagonal gauge fields it is natural to ask whether it is
possible to a) remove the mass terms as explicit infrared regulators,
b) maintain finite mass terms for the off-diagonal gauge fields in the
effective action, but c) recover the standard Slavnov-Taylor identities
without additional terms. We find, within a parametrization of the
effective action by its perturbatively relevant terms only, that this
situation is possible. Of course, a sufficiently violent (divergent)
infrared behaviour of the wave function renormalization constants is
required for this purpose. Most interestingly the Slavnov-Taylor 
identities imply that, amongst others, it is the wave function
renormalization constant of the diagonal gauge fields (which are
supposedly responsable for confinement and vanishing in the infrared
\cite{18}), which appears in the denominator of the renormalised mass
term for the off-diagonal gauge fields and could thus render it
non-vanishing even in the limit of vanishing bare mass term. Of course
we cannot prove this behaviour here. Although this scenario would allow
for standard Slavnov-Taylor identities together with massive
off-diagonal gauge fields, this is not sufficient for perturbative
unitarity. On the other hand, perturbative unitarity in a confining
theory is possibly not even required; thus we will not discuss
perturbative unitarity any further in the rest of the paper.\par

The paper is organized as follows. First, in section 2, we introduce
the model (massive Yang-Mills theory in an abelian gauge) and its 
symmetries. In section 3 we prove its renormalisability. For
simplicity, the proofs are presented for SU(2), but the generalisation
to SU(N) or the inclusion of gauge invariant matter is straightforward.
The massless limit is smooth and the usual Slavnov-Taylor identities
are recovered in this limit. In section 4 we study first the CS
equations. We discuss some properties of the $\beta$ functions and
anomalous dimensions, and the different roles of the masses for the
diagonal and off-diagonal gluons. We comment the relation with the ERG
equations in this respect, and finish with a speculation on the
possibility to obtain massive off-diagonal gluons in the confining
phase. Conclusions are drawn in section 5.

\mysection{The Model}
\hspace*{\parindent} 
 
We consider pure $SU(2)$ Euclidean Yang-Mills theory. $A_\mu$ denote
the gauge fields associated to the $U(1)$ subgroup, and $W_{\mu}^\pm$
are the off-diagonal gauge fields with the corresponding $U(1)$
charges. The classical action $S$ reads
 
\beq
\label{2.1e}
 S=\int d^4x\left\lbrace {\mathcal{L}} _{YM}+{\mathcal{L}} _{GF}+ 
 {\mathcal{L}}_{m}\right\rbrace
\eeq

\noi where ${\mathcal{L}} _{YM}$ is the Yang-Mills Lagrangian

\bea
\label{2.2e}
{\mathcal{L}} _{YM} &=& 
\frac{1}{4}(\partial_{\mu}A_{\nu}-\partial_{\nu}A_{\mu})^2
+\frac{ig}{2}(\partial_{\mu}A_{\nu}
-\partial_{\nu}A_{\mu})(W_{\mu}^+W_{\nu}^--W_{\mu}^-W_{\nu}^+)
\nn \\
&& -\frac{g^2}{4}(W_{\mu}^+W_{\nu}^-
-W_{\mu}^-W_{\nu}^+)^2 
+\frac{1}{2}(D_{\mu}W_{\nu}^+
-D_{\nu}W_{\mu}^+)(D_{\mu}W_{\nu}^-
-D_{\nu}W_{\mu}^-)\ , \nn \\
\eea

\noi and ${\mathcal{L}}_{GF}$ is the gauge fixing part

\bea
\label{2.3e}
{\mathcal{L}}_{GF}&=&-\frac{\beta}{2}\phi_{3}^{2}-\phi_{3}\partial_{\mu}
A_{\mu}+\partial_{\mu}\overline{c}^{3}(\partial_{\mu}c^{3}
+ig(W_{\mu}^{+}c^{-}-W_{\mu}^{-}c^{+}))\nn \\
&&-\alpha(\phi^{+}-igc^{3}\overline{c}^{+})(\phi^{-}+igc^{3}
\overline{c}^{-})\nn \\
&& -(\phi^{+}-igc^{3}\overline{c}^{+})D_{\mu}W_{\mu}^{-}
-(\phi^{-}+igc^{3}\overline{c}^{-})D_{\mu}W_{\mu}^{+}
+D_{\mu}\overline{c}^{+}D_{\mu}c^{-}+D_{\mu}
\overline{c}^{-}D_{\mu}c^{+} \nn \\
&&+g^2(W_{\mu}^{+}\overline{c}^{-}-W_{\mu}^{-}\overline{c}^{+})
(W_{\mu}^{+}c^{-}-W_{\mu}^{-}c^{+})
-g^2\alpha \overline{c}^{+} c^{-} \overline{c}^{-}c^{+}\ . \nn \\
\eea

\noi The derivatives $D_\mu$ are $U(1)$ covariant. For convenience we
write the masses for the gauge bosons  $W_\mu^\pm$, $A_\mu$, charged
and neutral ghosts in the form $m^2$, $\lambda m^2$, $\chi_c m^2$ and
$\chi_3 m^2$, respectively:

\bea
\label{2.4e}
{\mathcal{L}}_{m}&=&m^2(W_{\mu}^+W_{\mu}^{-}+\chi_c 
(\overline{c}^{+}c^{-}+\overline{c}^{-}c^{+})
+\frac{\lambda}{2}A_{\mu}A_{\mu}+\chi_3 \overline{c}^3c^3)
\eea

In the massless case, renormalisability  for a two parameter family of
abelian gauges has been shown in \cite{19}. Our choice for
${\mathcal{L}}_{GF}$ corresponds to a particular choice for the
parameters in \cite{19}, which has also been studied and proven to be
stable under renormalization in \cite{20}. The techniques used are the
usual BRST \cite{21} symmetry and some additional identities exposed
below. \par

${\mathcal{L}}_{GF}$ is invariant under the (nilpotent) BRST
transformations

$$\begin{array}{lll}
&sA_{\mu}=\partial_{\mu}c^3\pm igW_{\mu}^{\pm}c^{\mp},\qquad
&sW_{\mu}^{\pm}=D_{\mu}c^{\pm}\mp igc^3W_{\mu}^{\pm}, 
\nn \\
&sc^3=-igc^+c^-,          &sc^{\pm}=\mp igc^3c^{\pm},
\nn \\
&s\overline{c}^3=-\phi_3,  &s\overline{c}^\pm=-\phi^\pm,
\nn \\
&s\phi_3=0,               &s\phi^{\pm}=0.  \\
\end{array}$$ 
\beq\label{2.5e} \eeq

\noi Here $s$ denotes, as usual, the operator generating BRST
transformations:

\beq \label{2.6e}
\delta \varphi =\xi s \varphi
\eeq

\noi where $\xi$ is a Grassmann parameter. (Here and below $\varphi$
denote all possible fields; whenever necessary, we indicate explicitely
its $U(1)$ charge $\pm 1$.) We introduce the usual sources for
composite operators for the BRST variations and, in addition, a source
$\rho$ for the variation of ${\mathcal{L}}_{m}$:

\beq \label{2.7e}
{\mathcal{L}}\longrightarrow {\mathcal{L}}-K_{\mu}^3sA_{\mu}-
K_{\mu}^{\pm}sW_{\mu}^{\mp}-L^3sc^3-L^{\pm}sc^{\mp} - \rho
\Sigma_{\rho}
\eeq

\noi with

\bea
\label{2.8e}
\Sigma_{\rho}&=& s(W_{\mu}^+W_{\mu}^{-}+\chi_c (\overline{c}^{+}c^{-}+
\overline{c}^{-}c^{+})+\frac{\lambda}{2}A_{\mu}A_{\mu}+\chi_3
\overline{c}^3c^3)
\nn \\
&=& W_{\mu}^{\mp}D_{\mu}c^{\pm}
-\chi_c(\phi^{\pm}\mp igc^3\overline{c}^{\pm})c^{\mp} 
+\lambda A_{\mu}(\partial_{\mu}c^3\pm igW_{\mu}^{\pm}c^{\mp}) \nn \\
&&+\chi_3(ig\overline{c}^3c^+c^--\phi_3c^3)
\eea

\noi Now the mass terms lead to a relatively simple modification of the
Slavnov-Taylor identity for the action, which becomes

\bea \label{2.9e} 
&&\int d^{4}x \left\lbrace \frac{\delta \Gamma}{\delta K_{\mu}^3}
\frac{\delta \Gamma}{\delta A_{\mu}}
+\frac{\delta \Gamma}{\delta K_{\mu}^{\pm}} \frac{\delta \Gamma}{\delta
W_{\mu}^{\mp}}
+\frac{\delta \Gamma}{\delta L^3} \frac{\delta \Gamma}{\delta c^3}
+\frac{\delta \Gamma}{\delta L^{\pm}} \frac{\delta \Gamma}{\delta
c^{\mp}} -\phi_3 \frac{\delta \Gamma}{\delta \overline{c}^3}
-\phi^{\pm} \frac{\delta \Gamma}{\delta
\overline{c}^{\mp}}\right\rbrace \nn \\
&&+m^2\frac{\partial \Gamma}{\partial
\rho}=0 \ .
\eea

In addition to the non-linear Slavnov-Taylor identity (2.9) the action
has the following linearly realised symmetries: 

a) ghost number conservation induced by $\delta c=\epsilon c, \delta
\overline{c}=-\epsilon \overline{c}$, 

b) charge conjugation $\varphi^3\rightarrow -\varphi^3, \varphi^{\pm} 
\rightarrow \varphi^{\mp}$,

c) invariance under a constant shift of the neutral antighost, which 
is only broken by the neutral ghost mass and a term proportional to 
$\rho$, which can be written as a variation with respect to $L^3$.
It gives rise to the identity

\bea
\label{2.10e} 
\frac{\delta \Gamma}{\delta \overline{c}^{3}}
- \partial_{\mu}\frac{\delta \Gamma}{\delta K_{\mu}^3}-m^2 \chi_3
c^3-\chi_3 \rho \frac{\delta \Gamma}{\delta L^3}=0 \ .
\eea

d) the U(1) gauge symmetry is only broken by the gauge-fixing term
$\phi_{3}\partial_{\mu}A_{\mu}$, the $A_{\mu}$ mass term and a term 
proportional to $\rho$, which can be expressed in terms of a variation
with respect to $K^3_\mu$. An infinitesimal $U(1)$ transformation
allows to derive the following Ward identitiy:

\beq \label{2.11e}  
\partial_{\mu}\frac{\delta \Gamma}{\delta A_{\mu}}
\pm ig \varphi^{\pm}\frac{\delta \Gamma}{\delta \varphi^{\pm}}
-\partial^2\phi_3
-\rho\partial_{\mu}\frac{\delta \Gamma}{\delta K_{\mu}^3} 
-m^2\lambda \partial_{\mu}A_{\mu}=0 
\eeq

e) the Grassmannian U(1) gauge symmetry 

\beq 
\label{2.12e} 
\delta c^3(x)=\eta(x),\ 
\delta \phi^{\pm}(x)=\pm ig\eta (x) \overline{c}^{\pm}(x)
\eeq 

\noi is only broken by the neutral ghost kinetic term and terms
proportional to the sources $K$, $L$ and $\rho$, which gives raise to
the identity:

\bea
\label{2.13e}
\frac{\delta \Gamma}{\delta c^{3}}
\pm ig \overline{c}^\pm\frac{\delta \Gamma}{\delta \phi^{\pm}}
&=&(\partial^2-m^2\chi_3)\overline{c}^3
-\partial_{\mu}K_{\mu}^3
+ ig(\pm K_{\mu}^\pm W_{\mu}^\mp \mp L^\pm c^\mp ) \nn \\
&&-\rho \lambda \partial_{\mu}A_{\mu}-\rho \chi_3 \phi_3\ .
\eea
\noi In \cite{20} this equation (for $\lambda = \chi_3 = 0$) has been
denoted "neutral ghost field equation". It implies that the charged
fields $\phi^{\pm}$ and the neutral ghosts appear in the effective
action only in the combination $\phi^{\pm} \mp
igc^{3}\overline{c}^{\pm}$, apart from the terms linear in the other
fields which will thus remain unrenormalised.

\mysection{Proof of perturbative renormalisability}
\hspace*{\parindent} 

The proof of perturbative renormalisability requires an ultraviolet 
regulator (like dimensional regularisation) which respects all the
above symmetries or identities, and goes by induction in the number of
loops. Using usual power-counting arguments, new infinities to loop
order $n$ can appear only proportional to terms in the effective action
$\Gamma$ with dimension less or equal than four, once all infinities to
all loop orders less than $n$ have been reabsorbed into redefinitions
of couplings and fields. One has to show that all possible new
divergencies can be reabsorbed into redefinitions of the fields,
sources and parameters already present.

First we note that no divergencies quadratic in the sources $K$, $L$
and $\rho$ are possible. The following facts are required to this end:
$K$, $L$ and $\rho$ have mass dimensions 2, 2 and 1, respectively. $K$
and $\rho$ anticommute and have ghost number $-1$. The sources $L$ are
bosonic and have ghost number $-2$. Then ghost number conservation
implies that no terms of dimension four or less exist with two or more
sources $K$ and/or $L$, since one would need too many additional ghost
fields in order to end up with ghost number $0$. The same argument 
forbid terms $\sim K\cdot L$ and $\sim K\cdot \rho$. In addition,
because $\rho$ anticommutes, terms without derivatives $\sim\rho^2$
vanish. Hence we cannot construct terms quadratic in any sources with
vanishing ghost number and dimension less or equal than four.

The (divergent) contributions to $\Gamma$ linear in the sources
$K^{\pm}_{\mu}$, $K^{3}_{\mu}$, $L^{\pm}$ and $L^3$, to loop order
$n$, modify the effective BRST variations of the fields
$W^{\pm}_{\mu}$, $A_{\mu}$, $c^{\pm}$ and $c^3$ to loop order $n$. They
are strongly constraint by the Slavnov-Taylor identity (2.9), and the
constraints a) -- e) on the effective action. Let us denote the
effective BRST variations to loop order $n$ by

\beq
\label{3.1e}
s_n\varphi = s\varphi + \Delta {\widetilde s_n}\varphi
\eeq

\noi where $s\varphi$ is of the classical form (2.3) by assumption.
$\Delta {\widetilde s_n}\varphi$ is the divergent contribution from 
loop order $n$, which can be obtained from the coefficient of the
corresponding source terms. We have $\Delta {\widetilde s_n}
\overline{c} = 0$ (since, from (2.9),  $s_n \overline{c} = s
\overline{c} = -\phi$ to all loop orders) and $\Delta {\widetilde s_n}
\phi = 0$ (since, from (2.9), $s_n \phi = s \phi = 0$ to all loop
orders). The terms linear in $K$ and $L$ in the Slavnov-Taylor identity
(2.9) (where the term $\sim m^2$ does not contribute) to order $\Delta$
imply 

\beq
\label{3.2e}
s_n^2\varphi = 0+ {\cal O}(\Delta^2)
\eeq

\noi on the fields $A_{\mu}$ $W_{\mu}$ and $c$, which means that the
quantum corrected BRST variations remain nilpotent. Taking the
constraints a) -- e) on the effective action into account, the most
general form of the BRST variations to order $n$ is then

$$\begin{array}{lll}
&s_nA_{\mu}=\partial_{\mu}c^3+ig\sqrt{\frac{Z_WZ_cY_3}{Z}}
(W_{\mu}^+c^--W_{\mu}^-c^+),
&s_nc^3=-igY_3 Z_cc^+c^-,\\
&s_nW_{\mu}^{\pm}=\sqrt{\frac{ZY_3Z_c}{Z_W}}D_{\mu}c^{\pm}\mp
igc^3W_{\mu}^{\pm},
&s_nc^{\pm}=\mp igc^3c^{\pm},\\
&s_n\overline{c}^3=-\phi_3,
&s_n\overline{c}^\pm=-\phi^\pm,\\
&s_n\phi_3=0,
&s_n\phi^{\pm}=0.
\end{array}
$$
\beq \label{3.3e}\eeq

\noi The divergent constants $Z$, $Z_W$, and $Z_c$ will play the role
of the wave function renormalization constants of the fields $A_{\mu}$,
$W_{\mu}$ and $c^{\pm}$, and $\sqrt{Y_3}$ resp. $\sqrt{Y_3^{-1}}$ will
renormalize the neutral ghosts $c^3$ resp. $\overline{c}^3$. The
non-renormalization of the terms $\sim c^3$ in the variations  has its
origin in the identity (2.13).

Next we consider the quantum corrections to $\Sigma_\rho$, the BRST 
variation of the
mass term. Let us denote it by

\beq
\label{3.4e}
\Sigma_{n,\rho}=\Sigma_{\rho}+\Delta{\widetilde \Sigma_{n,\rho}}
\eeq

\noi where the first term on the right hand side is of the classical
form (2.8) by assumption, and $\Delta {\widetilde \Sigma_{n,\rho}}$ is
the divergent contribution from loop order $n$. 

The terms of order $\rho$ in the Slavnov-Taylor identity (2.9) now
imply (again the term $\sim m^2$ does not contribute)

\beq
\label{3.5e}
s_n \Sigma_{n,\rho} = 0 + {\cal O}(\Delta^2).
\eeq

\noi Together with the constraints a) -- e) $\Sigma_{n,\rho}$  
is then necessarily of the form

\bea
\label{3.6e}
&&\Sigma_{n,\rho}=\lambda A_{\mu}
\left(\partial_{\mu} c^3
+ig\sqrt{\frac{Y_3Z_WZ_c}{Z}}(W_{\mu}^+c^--W_{\mu}^-c^+)\right)\nn \\
&&
+Z_m\left(W_{\mu}^{\pm}(\sqrt{Z Y_3 Z_W Z_c}D_{\mu}c^{\mp}\pm
ig\sqrt{Z_W}c^3W_{\mu}^{\mp})
-Z_c Z_{\chi}\chi_c(\phi^{\pm}\mp
gc^3\overline{c}^{\pm})c^{\mp} \right).\nn \\
&&+\chi_3 \left(-\phi_3c^3+igY_3Z_c \overline{c}^3c^+c^- \right) \quad
\eea

The terms of zero order in $K$, $L$ and $\rho$ in (2.9) finally give
\beq
\label{3.7e}
s_n \Gamma_{n} = m^2 \Sigma_{n,\rho} + {\cal O}(\Delta^2).
\eeq

\noi where we have defined $\Gamma_n = S + \Delta\Gamma_n$. (Again $S$
is the classical action, and $\Delta\Gamma_n$ is the divergent
contribution from loop order $n$ to the source independent
part of the effective action.) It implies that also $\Gamma_n$ is
BRST invariant up to the mass terms.  The most general parametrization
of $\Gamma_{n}$ is then

\beq
\label{3.8e}
\Gamma_{n}=\Gamma_{n,YM}+\Gamma_{n,GF}+\Gamma_{n,m}
\eeq

\noi with

\bea
\label{3.9e}
\Gamma_{n,YM}&=&ZS_{YM}\left\lbrack A,W\sqrt{\frac{Z_W}{Z}}
\right\rbrack=\int d^4x \left\lbrace
\frac{Z}{4}(\partial_{\mu}A_{\nu}-\partial_{\nu}A_{\mu})^2\right.\nn \\
&&\left. +\frac{ig}{2}Z_W(\partial_{\mu}A_{\nu}
-\partial_{\nu}A_{\mu})(W_{\mu}^+W_{\nu}^--W_{\mu}^-W_{\nu}^+)
\right.\nn \\
&&\left. -\frac{g^2}{4}\frac{Z_W^2}{Z}(W_{\mu}^+W_{\nu}^-
-W_{\mu}^-W_{\nu}^+)^2\right.\nn \\
&&\left. +\frac{Z_W}{2}(D_{\mu}W_{\nu}^+
-D_{\nu}W_{\mu}^+)(D_{\mu}W_{\nu}^-
-D_{\nu}W_{\mu}^-)\right\rbrace\nn \\
\Gamma_{n,GF}&=&\int d^4x \left\lbrace -\frac{\beta}{2}\phi_{3}^{2}
-\phi_{3}\partial_{\mu}A_{\mu}
+\partial_{\mu}\overline{c}^{3}(\partial_{\mu}c^{3}
+ig\sqrt{\frac{Z_WZ_cY_3}{Z}}(W_{\mu}^{+}c^{-}-W_{\mu}^{-}c^{+}))
\right.
\nn \\
&&\left. -\sqrt{\frac{Z_WZ_c}{ZY_3}}(\phi^{+}
-igc^{3}\overline{c}^{+})D_{\mu}W_{\mu}^{-}
-\sqrt{\frac{Z_W Z_c}{Z Y_3}}(\phi^{-}
+igc^{3}\overline{c}^{-})D_{\mu}W_{\mu}^{+}\right.\nn \\
&&\left. +Z_cD_{\mu}\overline{c}^{+}D_{\mu}c^{-}
+Z_cD_{\mu}\overline{c}^{-}D_{\mu}c^{+}\right.\nn \\
&&\left. +g^2\frac{Z_WZ_c}{Z}(W_{\mu}^{+}\overline{c}^{-}
-W_{\mu}^{-}\overline{c}^{+})
(W_{\mu}^{+}c^{-}-W_{\mu}^{-}c^{+})\right.\nn \\
&&\left. -\alpha g^2 \frac{Z_{\alpha}Z_c^2}{Z}\overline{c}^{+} c^{-}
\overline{c}^{-}c^{+}
-\alpha \frac{Z_{\alpha}Z_c}{ZY_3}(\phi^{+}-igc^{3}\overline{c}^{+})
(\phi^{-}+igc^{3}\overline{c}^{-})\right\rbrace\nn \\
\Gamma_{n,m}&=& m^2\int d^4x \left\lbrace
\frac{\lambda}{2} A_{\mu}A_{\mu}
+Z_m Z_W W_{\mu}^+W_{\mu}^- 
\right. \nn \\ &&\left. 
+\chi_c Z_m Z_{\chi}Z_c (\overline{c}^+c^-+\overline{c}^-c^+)
+\chi_3 \overline{c}^3c^3
\right\rbrace
\eea

The mass terms $\lambda m^2 A_{\mu}A_{\mu}$ and $\chi_3 m^2 
\overline{c}^3 c^3$ remain unrenormalized thanks to the identities 
(2.10) and (2.11), whereas the  $W_{\mu}^{\pm}$ and charged ghost
masses require in general independent renormalizations denoted by $Z_m$
and $Z_m Z_{\chi}$, respectively. In addition to the wave function
renormalization constants $Z, Z_W, Z_c, Y_3$ we have introduced a gauge
parameter renormalization constant $Z_{\alpha}$, whereas no independent
renormalization of $\beta$ is allowed. 

Note that the quantum effective action (3.9) (including the source terms
$\sim$ $K$, $L$ and $\rho$ given implicitely by eqs. (3.3) and (3.6)) is
not the renormalized action, but just describes in a compact way the
possible UV divergencies at loop order $n$. In order to renormalize the
theory, counter terms have to be added to the bare Lagrangian such that
the effective action (3.9) becomes finite. This can be achieved by the
following rescaling of the  fields, sources and parameters:

$$ \begin{array}{llll}
\medskip
&A_{\mu,R}=\sqrt{Z}A_{\mu},
&c_{R}^{3}=\frac{1}{\sqrt{Y_3}}c^{3},
&\overline{c}_{R}^{3}=\sqrt{Y_3}\overline{c}^{3},\nn \\
\medskip
&W_{\mu,R}^{\pm}=\sqrt{Z_W}W_{\mu}^{\pm},
&c_{R}^{\pm}=\sqrt{Z_c}c^{\pm},
&\overline{c}_{R}^{\pm}=\sqrt{Z_c}\overline{c}^{\pm},\nn \\
\medskip
&\phi_{R}^{3}=\frac{1}{\sqrt{Z}}\phi_{3},
&\phi_{R}^{\pm}=\sqrt{\frac{Z_c}{ZY_3}}\phi^{\pm},
&\nn \\
\medskip
&K_{\mu,R}^{3}=\sqrt{Y_3}K_{\mu}^3,
&K_{\mu,R}^{\pm}=\sqrt{\frac{ZY_3}{Z_W}}K_{\mu}^{\pm},
&\nn \\
\medskip
&L_{R}^{3}=Y_3\sqrt{Z}L^{3},
&L_{R}^{\pm}=\sqrt{\frac{Y_3Z}{Z_c}}L^{\pm},
&\nn \\
\medskip
&m_R^2=Z_mm^2,
&\rho_R=Z_m\sqrt{ZY_3}\rho,&\nn \\
&\lambda_R=\frac{\lambda}{Z_mZ},
&\chi_{c,R}=Z_{\chi} \chi_c, 
&\chi_{3,R}=\frac{\chi_3}{Z_m},\nn \\
\medskip
&g_R=\frac{1}{\sqrt{Z}}g,
&{\alpha}_R=Z_{\alpha}\alpha,
&\beta_R=Z \beta.
\end{array} $$ 
\beq \label{3.10e} \eeq

Thus the model is perturbatively renormalisable. Note that in our
convention Green functions of the renormalized fields $A_{\mu, R}$ etc.
are finite (as is the quantum effective action); the wave function
renormalization constants $Z$ etc. correspond to $Z^{-1}$ etc. used
frequently in the literature \cite{23}. (Our factors $Z$ describe the UV
divergencies to loop order $n$ rather than the required counter terms,
which are proportional to $Z^{-1}$.)

With a gauge invariant UV regulator (as dimensional regularization) the
renormalization of all mass terms is multiplicative; quadratically
divergent additive renormalization constants would only be required if
one employs a naive UV cutoff. In dimensional regularization, the
renormalisation constants can be chosen independent from the masses
\cite{22}. 

For completeness we give here the one loop expressions for all
renormalization constants:

\bea \label{3.11e}
&Z=&1-g^2\frac{22}{3} I \nn \\
&Z_c=&1-g^2 (3-\beta) I\nn \\
&Z_W=&1-g^2\left(\frac{22}{3} - \frac{9+\alpha}{2} -\beta\right) I\nn \\
&Y_3 =& 1+g^2(3+\alpha) I\nn \\
&Z_{\alpha}=&1-g^2\left( \frac{4}{3} - \alpha  -\frac{3}{\alpha}
\right) I\nn \\
&Z_{\chi} =& 1-g^2\left( \frac{4}{3}+2\alpha +\beta -2\chi_c -\frac{3}
{\chi_c} -\frac{\alpha^2}{\chi_c}\right. \nn \\
&&\left. +\lambda \left(\frac{9+3\alpha}{4} +\frac{3}{\chi_c}
+\beta(\beta -1) (\frac{3}{2} +\frac{1}{2\alpha})\right)\right) I\nn \\
&Z_m =& 1-g^2\left( -\frac{13}{3} - \alpha +2\chi_c
+\lambda\left(-\frac{9+3\alpha}{4} +\beta(1-\beta)(\frac{3}{2}
+\frac{1}{2\alpha})\right) \right) I\nn \\
&&
\eea

\noi where $I$ denotes the ultraviolet divergent part of the
logarithmically divergent integral

$$\begin{array}{lll}
I=\left[\int_{reg}\frac{d^d p}{(2\pi)^d}\frac{1}{p^4}\right]_{div.}
&= \frac{1}{16\pi^2\epsilon}\quad 
&\rm{in}\ d=4-\epsilon\ \rm{dimensions} \\
&= \frac{1}{16\pi^2}\ln{\Lambda^2}\quad 
&\rm{in\ d=4\ with\ an\ UV\ cutoff}\ \Lambda .
\end{array}$$
\beq \label{3.12e}\eeq

The classical (or renormalized) action of the model can actually be
simplified considerably, if we integrate out the auxiliary fields
$\phi^{\pm}$, i.e. replace them by the solutions of their (algebraic)
equations of motion. Then the classical gauge fixing part of the
Lagrangian in (2.1) reads

\bea
\label{3.13e}
{\mathcal{L}}_{GF}^*&=&-\frac{\beta}{2}\phi_{3}^{2}
-\phi_{3}\partial_{\mu}A_{\mu}
+\partial_{\mu}\overline{c}^{3}(\partial_{\mu}c^{3}
+ig(W_{\mu}^{+}c^{-}-W_{\mu}^{-}c^{+}))
+\frac{1}{\alpha}D_{\mu}W_{\mu}^{-}D_{\nu}W_{\nu}^{+}\nn \\
&&+D_{\mu}\overline{c}^{+}D_{\mu}c^{-}
+D_{\mu}\overline{c}^{-}D_{\mu}c^{+}
+g^2(W_{\mu}^{+}\overline{c}^{-}-W_{\mu}^{-}\overline{c}^{+})
(W_{\mu}^{+}c^{-}-W_{\mu}^{-}c^{+})\nn \\
&&-g^2\alpha \overline{c}^{+} c^{-} \overline{c}^{-}c^{+} \ .
\eea

\noi Apart from the fixing of the abelian gauge group, this is the
continuum limit of the non-perturbative gauge fixing prescription
proposed by Schaden \cite{12,7}. Now, in addition, one could integrate
out the neutral ghosts and Lagrange multiplier $\phi_3$ preserving a
local action (provided the determinant of the Laplacian is correctly
regularised). It is interesting to note that, apart from the mass
terms, the classical action is now invariant under two independent BRST
symmetries:

$$\begin{array}{lll}
&s_1A_{\mu}=ig(W_{\mu}^+c^--W_{\mu}^-c^+)
&s_2A_{\mu}=\partial_{\mu} c^3\\
&s_1W_{\mu}^{\pm}=D_{\mu}c^{\pm}
&s_2W_{\mu}^{\pm}=\mp igc^3W_{\mu}^{\pm}\\
&s_1c^{\pm}=0
&s_2c^{\pm}=\mp igc^3c^{\pm}\\
&s_1c^3=-igc^+c^-
&s_2c^3=0\\
&s_1\overline{c}^{\pm}=\frac{1}{\alpha}D_{\mu}W_{\mu}^{\pm}
&s_2\overline{c}^{\pm}=\mp igc^3\overline{c}^{\pm}\\
&s_1\overline{c}^3=-\phi_3
&s_2\overline{c}^3=-\phi_3\\
&s_1\phi_3=0
&s_2\phi_3=0
\end{array} $$
\beq \label{3.14e} \eeq

The second symmetry is just a particular $U(1)$ gauge transformation
with gauge parameter $\xi c^3$ and is nilpotent. The first symmetry
$s_1$ is not nilpotent, but $s_1^2$ gives again just a $U(1)$ gauge
transformation \cite{12,7}. 

The divergent part in the effective action in the formulation without
lagrange multipliers can be obtained eliminating them in (3.9). The 
gauge-fixing part then reads:

\bea
\label{3.15e}
\Gamma_{n,GF}^{*}
&=&\int d^4x \left\lbrace -\frac{\beta}{2}\phi_{3}^{2}
-\phi_{3}\partial_{\mu}A_{\mu}
+\partial_{\mu}\overline{c}^{3}(\partial_{\mu}c^{3}
+ig\sqrt{\frac{Z_WZ_cY_3}{Z}}
(W_{\mu}^{+}c^{-}-W_{\mu}^{-}c^{+}))\right.
\nn \\
&&\left. +Z_cD_{\mu}\overline{c}^{+}D_{\mu}c^{-}
+Z_cD_{\mu}\overline{c}^{-}D_{\mu}c^{+}\right. \nn \\
&&\left. 
+g^2\frac{Z_WZ_c}{Z}(W_{\mu}^{+}\overline{c}^{-}
-W_{\mu}^{-}\overline{c}^{+})
(W_{\mu}^{+}c^{-}-W_{\mu}^{-}c^{+})\right. \nn \\
&&\left. -\alpha g^2 \frac{Z_{\alpha}Z_c^2}{Z}\overline{c}^{+} c^{-}
\overline{c}^{-}c^{+}+\frac{Z_W}{\alpha
Z_{\alpha}}D_{\mu}W_{\mu}^+D_{\nu}W_{\nu}^- \right\rbrace\ .
\eea

\noi 
The model could be formulated without auxiliary fields from the
beginning, and the result can be proved to be the same, in spite of the
non-nilpotency of the first BRST variations $s_1$ in (3.14).

A particularly important case, which merits its own study, is where the
$A_{\mu}$ and $c^3$ mass terms $\lambda m^2 A_{\mu}A_{\mu}$ and $\chi_3
m^2 \overline{c^3}c^3$ vanish (by chosing $\lambda = \chi_3 = 0$). Now
it is convenient to chose $\chi_c$, the ratio of the $W_{\mu}^{\pm}$ to
$c^{\pm}$ masses, equal to the gauge parameter $\alpha$: Instead of
(2.4) we write for the $W_{\mu}^{\pm}$ and $c^{\pm}$ mass terms

\beq
\label{3.16e}
{\mathcal{L}}_{m}=m^2(W_{\mu}^+W_{\mu}^{-}+\alpha 
(\overline{c}^{+}c^{-}+ \overline{c}^{-}c^{+}))
\eeq

Let us return to the version with $\phi^{\pm}$ present: Comparing terms
$\sim \phi^{\pm}$ in the gauge fixing part (2.3) of the classical
Lagrangian to the term $\sim \rho$ in (2.7), i.e. the BRST variation of
the $W_{\mu}^{\pm}$ and $c^{\pm}$ mass terms, one finds that the
classical action satisfies an additional {\it linear} identity, which
remains thus unrenormalized:

\beq
\label{3.17e}
\frac{\partial\Gamma}{\partial\rho}=
-\int d^4x \lbrace c^{+}\frac{\delta\Gamma}{\delta\phi^{+}}
+c^{-}\frac{\delta\Gamma}{\delta\phi^{-}}\rbrace
\eeq

The source $\rho$ is now actually redundant, and the Slavnov-Taylor
identity (2.9) can be written without $\rho$: One can replace 
the term $\sim\ \partial\Gamma/\partial\rho$ by (3.17), which
corresponds to assigning a non-trivial BRST variation

\beq
\label{3.18e}
s\phi^{\pm}=-m^2c^{\pm}
\eeq

\noi instead of (2.5) to $\phi^{\pm}$. (Now, however, the BRST
variations on $\phi^{\pm}$ and the charged anti-ghosts are no longer
nilpotent.) Due to the identity (3.17) the renormalization constant
$Z_m$ of the mass term is no longer independent; now one has

\beq
\label{3.19e}
Z_m = (Z Y_3)^{-1}\ .
\eeq

\noi Instead of (3.9) the renormalized mass terms now read

\bmini
\label{3.20ae}
\Gamma_{n,m}&=& \int d^4x \frac{m^2}{ZY_3} \left\lbrace
Z_W W_{\mu}^+W_{\mu}^- 
+\alpha Z_{\alpha}Z_c (\overline{c}^+c^-+\overline{c}^-c^+)
\right\rbrace
\\
\label{3.20be}
&=& \int d^4x m_R^2 \left\lbrace
W_{R\mu}^+W_{R\mu}^- +
\alpha_R (\overline{c}^+_R c^-_R +\overline{c}^-_R c^+_R)
\right\rbrace
\emini

\noi Hence, after the redefinitions (3.10), the
classical form (3.16) (with its particular ratio among the 
$W_{\mu}^{\pm}$ and $c^{\pm}$ mass terms) remains preserved by 
renormalisation.

Note that it is quite non-trivial that mass terms for $W_\mu^\pm$ and
the charged ghosts {\it alone} allow to maintain quite simple 
Slavnov-Taylor identities, provided the BRST variation of the auxiliary
fields $\phi^\pm$ is modified (non-vanishing) according to eq. (3.18).

\mysection{Renormalization Group Flows}
\hspace*{\parindent} 

In massive theories the dependence of the Green functions on the 
masses can be decribed by the Callan-Symanzik (CS) equations. Likewise,
in the case of more general infrared cutoff terms quadratic in the
fields, the dependence on the infrared cutoff is given by the
Wilsonian Exact Renormalization Group (ERG) equations. In this section
we will discuss the features of these renormalization group flows in
Yang-Mills theories in abelian gauges.

Let us start with the CS equations. The renormalization program, i.e.
the determination of the counter terms order by order in perturbation
theory, requires a renormalization prescription. In the present case of
a massive theory a particularly simple renormalization prescription
exists: one can chose as many independent one-particle irreducible
Green functions as there are independent counter terms, and require
that they assume prescribed values at {\it vanishing} external momenta.
Typically one will chose two point functions and their second
derivatives with respect to the momentum (at vanishing momentum); these
are  particularly simple to evaluate. Now, in contrast to a minimal
substraction scheme, the counter terms will depend in a well defined
way on the masses.

Let us now assume that we have chosen a set of renormalized masses
$m_W^2 = m_R^2$, $m_A^2 = \lambda_R m_R^2$, $m_{c^3}^2 = \chi_{3R}
m_R^2$ and $m_{c^\pm}^2 = \chi_{cR} m_R^2$ with non-vanishing finite
parameters $\lambda_R$ and $\chi_R$ after renormalization. All Green
functions are ultraviolet finite by construction, as are the Green
functions using a slightly different renormalization prescription $m_R
\to m_R + \delta m_R$. Hence the derivatives of all Green functions
with respect to $m_R$ are equally ultraviolet finite.

In order to derive the CS equations one starts with the derivative of
the renormalized partition function with respect to the bare mass $m$.
This derivative hits both the mass terms in the bare action, as well as
the implicit mass dependence of all counter terms. The result can be
expressed in terms of the variation of the renormalized effective
action with respect to the renormalized mass. We will not rederive all
corresponding steps here, which are discussed in various textbooks
(see, e.g., \cite{23}). The CS equation for the renormalized effective
action then assumes the form

$$\left\lbrace
m_R \frac{\partial}{\partial m_R}
+\beta_g\frac{\partial}{\partial g_R}
+\beta_{\alpha}\frac{\partial}{\partial \alpha_R}
+\beta_{\beta}\frac{\partial}{\partial \beta_R}
+\beta_{\chi_c}\frac{\partial}{\partial \chi_{cR}}
+\beta_{\chi_3}\frac{\partial}{\partial \chi_{3R}}
+\beta_{\lambda}\frac{\partial}{\partial \lambda_R} 
\right. $$
\beq
\label{4.1e}
\left. +\sum_{i}\eta_i\int d^4x \varphi_i(x)
\frac{\delta}{\delta\varphi_i(x)}\right\rbrace \Gamma_R
= \Delta_m \Gamma_R \ .
\eeq

\noi Here we have introduced $\beta$ functions for the dimensionless
parameters of the theory, the gauge coupling $g$, the gauge parameters
$\alpha$ and $\beta$ (we hope that the reader will excuse this double
use of $\beta$) and the mass ratios $\lambda$ and $\chi$:

\bea
\label{4.2e}
&\beta_g=m_R\frac{\partial g_R}{\partial m_R}
&\beta_{\alpha}=m_R\frac{\partial \alpha_R}{\partial m_R}
\ \ \beta_{\beta}=m_R\frac{\partial \beta_R}{\partial m_R}\nn \\
&\beta_{\chi_c}=m_R\frac{\partial \chi_{cR}}{\partial m_R}
&\beta_{\chi_3}=m_R\frac{\partial \chi_{3R}}{\partial m_R}
\ \ \beta_{\lambda}=m_R\frac{\partial \lambda_R}{\partial m_R}\ .
\eea

\noi $\eta_i$ denote the various anomalous dimensions:

\beq
\label{4.3e}
\eta_i=\frac{m_R}{2}\frac{\partial log(Z_i)}{\partial m_R}\ .\qquad
\eeq

\noi The right hand side $\Delta_m \Gamma_R$ of eq. (4.1) consists of
the sum of (bare) mass terms and one loop diagrams with mass
insertions:

\bea
\label{4.4e}
\Delta_m \Gamma_R &=&
\frac{m_R^2\gamma_m}{Z_m}
\int d^dx\left\lbrace
Z_W^{-1}\left[ W_{\mu,R}^+W_{\mu,R}^-
+\frac{1}{2}Tr \left[\Gamma^{(2)}\right]_{W_{\mu,R}^+,W_{\nu,R}^-}^{-1}
\right]\right.\nn \\
&&\left. +\frac{\lambda_R Z_m}{2}Z_{\lambda}^{-1}\left[
A_{\mu,R}A_{\mu,R}+Tr\left[\Gamma^{(2)}\right]_
{A_{\mu,R},A_{\nu,R}}^{-1}\right] \right.\nn \\
&&\left. +\chi_{cR} Z_{\chi}^{-1}Z_c^{-1}\left[
\overline{c}_R^{\pm}c_R^{\mp}
-Tr\left[\Gamma^{(2)}\right]_{c_R^{\pm},\overline{c}_R^{\mp}}^{-1}
\right] \right.\nn \\
&&\left. +\chi_{3R} Z_c^{-1}\left[
\overline{c}^3_{R}c^3_{R}
-Tr\left[\Gamma^{(2)}\right]_{c^3_{R},\overline{c}^3_{R}}^{-1}
\right] 
\right\rbrace \qquad 
\eea

\noi where we have defined

\beq
\label{4.5e}
\gamma_m =\frac{Z_m}{m_R} \frac{\partial m^2}{\partial m_R}
= 2 - m_R  \frac{\partial log(Z_m)}{\partial m_R}\ .\qquad
\eeq

The expressions $\left[\Gamma^{(2)}\right]_{\varphi^a,\varphi^b}^{-1}$
denote the one loop diagrams which correspond to the $\varphi^a,
\varphi^b$ propagators (in an arbitrary background) at coincident
points. \par

Although the individual terms in (4.4) are not separately ultraviolet
finite, renormalizability ensures that their sums -- as well as all
$\beta$ functions and anomalous dimensions $\eta_i$ in (4.1) -- are
finite in the limit of an infinite ultraviolet cutoff $\Lambda$
\cite{23}.

The use of the CS equation (4.1) consists in the study of the behaviour
of Green functions at large Euclidean momenta $p^2$: After having
removed an ultraviolet cutoff $\Lambda$, dimensionless Green functions
can only depend on the ratio $p^2/m_R^2$ for dimensional reasons.
Furthermore, for $p^2/m_R^2 \gg 1$, the right hand side of (4.1)
becomes negligeable \cite{23}. Writing ${\partial}/{\partial m_R^2} =
-{\partial}/ {\partial p^2}$ the desired renormalization group equation
describing the $p^2$ dependence is then easily obtained.

In the case of Yang-Mills theories in abelian gauges one finds, using
our results of the preceeding section and in \cite{7,19,24}, that
several $\beta$ functions are related to the anomalous dimension
$\eta_A$ of $A_\mu$:

\bmini
\label{4.6ae}
\beta_g&=&-g_R\eta_A\\
\label{4.6be}
\beta_{\beta}&=&-2\beta_R\eta_A\\
\label{4.6ce}
\beta_{\lambda}&=&-\lambda_R(2-\gamma_m+2\eta_A)
\emini

\noi (Note that $\beta_R$ on the right hand side of (4.6b) denotes the 
renormalized gauge parameter.) 

Equation (4.6a) implies that $\beta_g$ can be computed from the $A_\mu$
propagator alone \cite{24}. Now recall eqs. (4.2), i.e. the fact that
in the context of the CS equations the $\beta$ functions are given by
the derivatives of the counter terms with respect to the masses. One 
finds that 

a) to one loop order, $A_\mu$ propagators do not contribute to the
renormalization of the $A_\mu$ propagator itself (i.e. $\eta_A$), and 

b) even to two loop order the renormalization of the $A_\mu$
propagator is infrared finite for $m_A\to 0$ or $\lambda \to 0$. 

This allows to obtain the two-loop $\beta$ function of Yang-Mills
theories in the particularly simple approach discussed near the end of
the preceeding section: It is possible to choose $m_A = m_{c^3} = 0$
(or $\lambda = \chi_3 = 0$) and $m_{c^\pm}^2 = \alpha m_W^2$ (or
$\chi_c =  \alpha$), and thus to invoke the particularily simple
version of the Slavnov-Taylor identities obtained by the use of eq.
(3.17). Up to the universal two loop order $\beta_g$ is then obtained
by the variation of $Z_A$ with respect to $m_W$ and $m_{c^\pm}$ only.

Eq. (4.6c) shows that $m_A = 0$ (or $\lambda = 0$) is stable under the
renormalization group flow, and furthermore eq. (4.5) together with
(3.19) implies that $\gamma_m$ is now given by

\beq
\label{4.7e}
\gamma_m=2(1+\eta_A+\eta_{\overline{c}^3})\ . \qquad
\eeq

However, for $m_A = 0$ not all Green functions are infrared finite at
vanishing external momenta. Already the one loop contributions to the
four point functions with four $W_\mu^{\pm}$ bosons (or two
$W_\mu^{\pm}$ bosons and two charged ghosts, or four charged ghosts)
involving two $W_\mu^{\pm} W_\mu^{\mp}A_\mu A_\mu$ vertices are
infrared divergent in this case. Hence non-vanishing external momenta
have to be chosen at the renormalization points for at least some
counter terms, and not all $\beta$ functions and anomalous dimensions
are given just by the variation with respect to $m_W$ (and
$m_c$) only.

The ERG equations \cite{16,17} describe equally the variation of the
Green functions with respect to an infrared cutoff, which is introduced
in the form of additional terms quadratic in the fields to the bare
action. In the case of Yang-Mill theories modified Slavnov-Taylor
identities have to be imposed \cite{17}; the corresponding
modifications correspond to the term $\sim \partial\Gamma/\partial
\rho$ in our eq. (2.9). (In the context of ERG equations a
corresponding source $\rho$ could, in principle, also be introduced;
this would not particularly simplify, however, a parametrization of the
effective action which satisfies the modified Slavnov-Taylor
identities.)

Various forms of such cutoff terms have been discussed in the
literature \cite{16,25}, and mass terms as discussed here would of
course be particularly simple. In the context of non-abelian Yang-Mills
theories they have been proposed in \cite{26}, but they lead to
conceptional problems. 

First, a mass term as infrared cutoff in the ERG context corresponds
to  a {\it bare} mass. Whereas the renormalizability of the theory
implies that derivatives of the Green function with respect to the {\it
renormalized} mass are ultraviolet finite (cf. the CS equations),
ultraviolet finiteness of  derivatives of Green function with respect
to the bare mass will generically not be given. For this reason of
ultraviolet finiteness of the ERG equations the infrared cutoff terms
have to have a non-trivial momentum dependence in general; notably a
simple mass term for $W_\mu^\pm$ and the charged ghosts only (with its
simple modifications of the  Slavnov-Taylor identities) can,
unfortunately, not be used to define an ultraviolet finite exact
renormalization group flow. Second, as discussed above, $m_W$ alone 
would not serve as an infrared regulator for all Green functions; the
remaining infrared divergences reduce to the ones of QED with matter
which are, however, much easier to deal with.

Let us add another twist to the difference between bare and
renormalized mass terms. Clearly, in order to recover the partition
function of QCD, we have to consider the limit of vanishing bare mass
terms. Is it possible to maintain non-vanishing renormalized masses in
this limit, and how can these agree with the Slavnov-Taylor
identities? This question is particularly acute in view of the finite
$W$ masses observed on the lattice \cite{3}.

First we note that our parametrization (3.9) of the quantum effective
action to $n$ loop order is not only useful in order to constrain the
corresponding ultraviolet divergences: After having renormalized the
theory, we can also ask what is the most general form (consistent with
the Slavnov-Taylor identities) of the perturbatively relevant terms in
the full quantum effective action. Again this is given by the
parametrization (3.9); now, however, the renormalization constants
$Z_i$ denote simply the behaviour of the corresponding Green functions
in the limit of vanishing external momenta: By definition, the
coefficient of each term (specified by its field content) in the
effective action denotes the one particle irreducible Green function
with the corresponding fields as (amputated) external lines. These
Green functions depend on the external momenta, and will approach the
form of the bare action in the limit of large Euclidean non-exceptional
external momenta. In the limit of vanishing external momenta, these
Green functions can well diverge in a massless theory.

Within this new interpretation of the parametrization (3.9) of the 
quantum effective action the parameters $g$, $m$, $\lambda$ and $\chi$ 
are finite (renormalized) quantities, and the coefficients $Z$, $Y$ are
ultraviolet finite numbers which possibly diverge for $m \to 0$.
Standard Yang-Mills theory is obtained in the limit $m \to 0$, since in
this limit the Slavnov-Taylor identities (2.9) turn into the standard
massless identities.

Let us now concentrate on the case $\lambda, \chi_3 = 0$, where the
mass terms for $A_\mu$ and the neutral ghosts have already been
switched off; then we are left to consider the mass terms $m^2 Z_m (Z_W
W_{\mu}^+W_{\mu}^-  +\chi_c Z_{\chi}Z_c
(\overline{c}^+c^-+\overline{c}^-c^+))$ in (3.9) in the limit $m^2 \to
0$. From (3.19) we find that they could actually remain finite in this
limit, provided the product $ZY_3$ vanishes sufficiently rapidly for
$m^2 \to 0$. This would allow for a massive behaviour of the $W$
propagator, maintaining the standard form of the Slavnov-Taylor
identities (together with standard BRST variations for $\phi^\pm$); 
however, one carefully has to study the effects of $ZY_3 \to 0$ in all
the other terms of the action:

a) For $Z \to 0$ the Slavnov-Taylor identities require the renormalized
coupling to diverge for vanishing external momenta. On the one hand  $Z
\to 0$, in $SU(N)$ covariant gauges, is believed to be a signal for
confinement \cite{18}, on the other hand lattice results \cite{27}
(again in $SU(N)$ covariant gauges) indicate that effective vertices
remain finite in this limit. It is not clear which of these results
holds in abelian gauges considered here; in fact the abelian model for
confinement proposed in \cite{10} would correspond to the case
$Z \to 0$.

b) $Y_3 \to 0$ seems to be a possibility specific to abelian gauges:
After the elimination of $\phi^\pm$ by its equations of motion, one
finds from eq. (3.15) that the effective action is well behaved in this
limit. (Actually, if one includes the counter terms in the BRST
variations (3.14) after the elimination of $\phi^\pm$, one finds that
all variations $s_1$ are proportional to $Y_3$ or $\sqrt{Y_3}$. Hence in
the limit $Y_3 \to 0$ the BRST symmetries reduce to the abelian ones,
which explains the possiblity to have finite $W$ masses consistent with 
standard Slavnov-Taylor identities, without infrared divergencies.) 

Note that, whereas $Z \to 0$ has an effect on the gluonic vertices of
the theory, $Y_3$ appears only in the gauge fixing part (and mass
terms) of the action; hence it resembles more to a gauge parameter.
Hence, {\it after} renormalization, one could envisage to tune both $m
\to 0$ and $Y_3 \to 0$ with $m^2/Y_3$ finite and to define a
renormalizable massive theory with standard Slavnov-Taylor identities
this way; (perturbative) unitarity of this model remains, however, to
be investigated.

\mysection{Conclusions}
\hspace*{\parindent} 

We have constructed a massive generalisation of SU(2)-Yang-Mills theory
in an abelian gauge. Its renormalizability can be shown thanks to a
simple generalisation of the Slavnov-Taylor identities. The
possibility to renormalise a (pure) Yang-Mills theory for fixed finite
masses (which can be chosen, due to asymptotic freedom, such that the
theory is completely perturbative) allows to separate the
renormalisation process from infrared phenomena. 

After completion of the renormalization program one will typically be
interested in removing the "artificial" mass terms. We have discussed,
in which cases this procedure defines consistent renormalization group
flows. Such renormalization group flows can actually be used directly
as tools in order to probe the infrared behaviour of Yang-Mills
theories. 

It is probable that confinement (via monopole condensation) is more
easily describable in abelian gauges. Then the massive version in such
gauges will be particularly useful. In the abelian gauge we have to
distinguish between the abelian "diagonal" and the "off-diagonal"
gluons, and we carefully discussed the different properties of the
different mass terms. We cannot identify these masses "for symplicity"
without spoiling the renormalisability. The possibility to maintain
finite renormalized masses for the (off-diagonal) $W$ gauge bosons for
vanishing bare mass terms seems to be promising, and may help to
understand some aspects of the infrared limit of Yang-Mills theories.

\vspace {1cm}

\noi {\Large \bf Acknowledgement}

\vspace {0.5cm}
We would like to thank L. Baulieu for helpful discussions.

\end{document}